\documentstyle[psfig]{mn0}

\def\simg{\mathrel{\rlap{\raise 0.511ex \hbox{$>$}}{\lower 0.511ex \hbox{$\sim$}}}}
\def\siml{\mathrel{\rlap{\raise 0.511ex \hbox{$<$}}{\lower 0.511ex \hbox{$\sim$}}}}
\def\ra{\hspace{-4mm}$\longrightarrow$} \def\xc{\hspace{-7mm}$\nu_x=\nu_c$}

\begin{document}

\title[Jet-Breaks in Swift Afterglows]
   {Jet-breaks in the X-ray Light-Curves of Swift GRB Afterglows }

\author[A. Panaitescu]{A. Panaitescu \\ Space Science and Applications, MS D466, 
                       Los Alamos National Laboratory, Los Alamos, NM 87545, USA}

\maketitle

\begin{abstract}
\small{
 In the set of 236 GRB afterglows observed by Swift between January 2005 and March 2007,
we identify 30 X-ray light-curves whose power-law fall-off exhibit a steepening ("break") 
at 0.1--10 day after trigger, to a decay steeper than $t^{-1.5}$. For most of these 
afterglows, the X-ray spectral slope and the decay indices before and after the break 
can be accommodated by the standard jet model although a different origin of the breaks
cannot be ruled out. In addition, there are 27 other afterglows whose X-ray light-curves 
may also exhibit a late break to a steep decay, but the evidence is not that compelling.
 The X-ray emissions of 38 afterglows decay slower than $t^{-1.5}$ until after 3 day,
half of them exhibiting such a slow decay until after 10 day. 
 Therefore, the fraction of well-monitored Swift afterglows with potential jet-breaks is 
around 60 percent, whether we count only the strongest cases for each type or all of them. 
This fraction is comparable to the 75 percent of pre-Swift afterglows whose optical 
light-curves displayed similar breaks at $\sim 1$ day.
 The properties of the prompt emission of Swift afterglows with light-curve breaks show 
the same correlations (peak energy of GRB spectrum with the burst isotropic output and 
with burst collimated output) as previously found for pre-Swift optical afterglows with 
light-curve breaks (the Amati and Ghirlanda relations, respectively). However, we find 
that Ghirlanda relation is largely a consequence of Amati's and that the use of the 
jet-break time leads to a stronger Ghirlanda correlation only when the few outliers to 
the Amati relation are included.
}
\end{abstract}


\section{Introduction}

 Observations of Gamma-Ray Burst (GRB) afterglows from 1999 to 2005 have evidenced the 
existence of breaks in the optical light-curve of many afterglows, occurring at 0.3--3 
day and being followed by a flux decay $F_o \propto t^{-\alpha}$, with the index 
$\alpha$ ranging from 1.3 to 2.8.
These breaks have been widely interpreted as due to the tight collimation of GRB outflows:
when the jet Lorentz factor decreases below the inverse of the jet half-opening (i.e. 
when the cone of relativistically beamed emission is wider than the jet), 
the observer "sees" the jet boundary, which leads to a faster decay of the afterglow
flux (synchrotron emission from the ambient medium shocked by the blast-wave).
This decay may be further "accelerated" by the lateral spreading of the jet (Rhoads 1999).

 The temporal--spectral properties of the afterglow optical emission are roughly 
consistent with the expectations of the standard jet model, if it assumed that 
$(i)$ the shock microphysical parameters and blast-wave kinetic energy are constant, 
and $(ii)$ the blast-wave kinetic energy per solid angle is the same along any direction 
This consistency yielded support to the jet interpretation for the optical light-curve 
breaks. Nevertheless, the basic confirmation of the jet model through observations of
achromatic breaks (i.e. exhibited by light-curves at different frequencies) lacked 
because of the limited coverage in the X-rays. To a large extent, because of the sparser 
optical follow-up, that proof is still modest today, despite the good X-ray monitoring 
of GRB afterglows provided by Swift.

 Collimation of GRB outflow is a desirable feature to decrease the burst output, as the 
largest isotropic-equivalent energy release approaches the equivalent of a solar mass 
(GRB 990123 -- Kulkarni et al 1999). From the light-curve break epoch of $\sim 1$ day, 
it follows that the half-opening $\theta_{jet}$ of GRB jets is of few/several degrees, 
for which the GRB output is reduced to $\siml 10^{51}$ erg (e.g. Frail et al 2001, 
Panaitescu \& Kumar 2001). However, such a tight collimation of the GRB outflow is more 
than necessary on energetic grounds, as the accretion of the debris torus formed during 
the collapse of a massive star (the origin of long-duration bursts -- e.g. Woosley 1993,
Paczy\'nski 1998) or of the black-hole spin can power relativistic jets with more than 
$10^{52}$ erg (e.g. M\'esz\'aros, Rees \& Wijers 1999, Narayan, Piran \& Kumar 2001). 
In other words, without demanding too much energy, GRB jets can be wider than several 
degrees, placing the afterglow jet-break epoch (which scales as $\theta_{jet}^{8/3}$ 
for a homogeneous circumburst medium and as $\theta_{jet}^4$ for a wind-like medium) 
later than usually reachable by afterglows observations.

 Recently, Burrows \& Racusin (2007) have suggested that Swift X-ray afterglows do not
display $\sim 1$ day breaks as often as pre-Swift optical afterglows. The purpose of
this article is 
$(i)$ to identify the afterglows observed by Swift from January 2005 through March 2007 
  whose X-ray light-curves steepen to a decay faster than $t^{-1.5}$ (which is, roughly, 
  the lower limit of the post-break decay indices of pre-Swift optical afterglows -- 
  figure 2 of Zeh, Klose \& Kann 2006),
$(ii)$ to find the Swift X-ray afterglows that were monitored longer than a few days 
  and displayed a decay slower than $t^{-1.5}$ (indicating a jet-break occurring well 
  after 1 day), and 
$(iii)$ to compare the fraction of Swift X-ray afterglows with potential jet-breaks 
  to that of pre-Swift optical afterglows with such breaks. 
In addition, the temporal--spectral properties of the X-ray afterglows will be used
to identify, on case-by-case basis, the required features of the standard jet model.

\section{Late breaks in X-ray afterglow light-curves}

\begin{figure}
\centerline{\psfig{figure=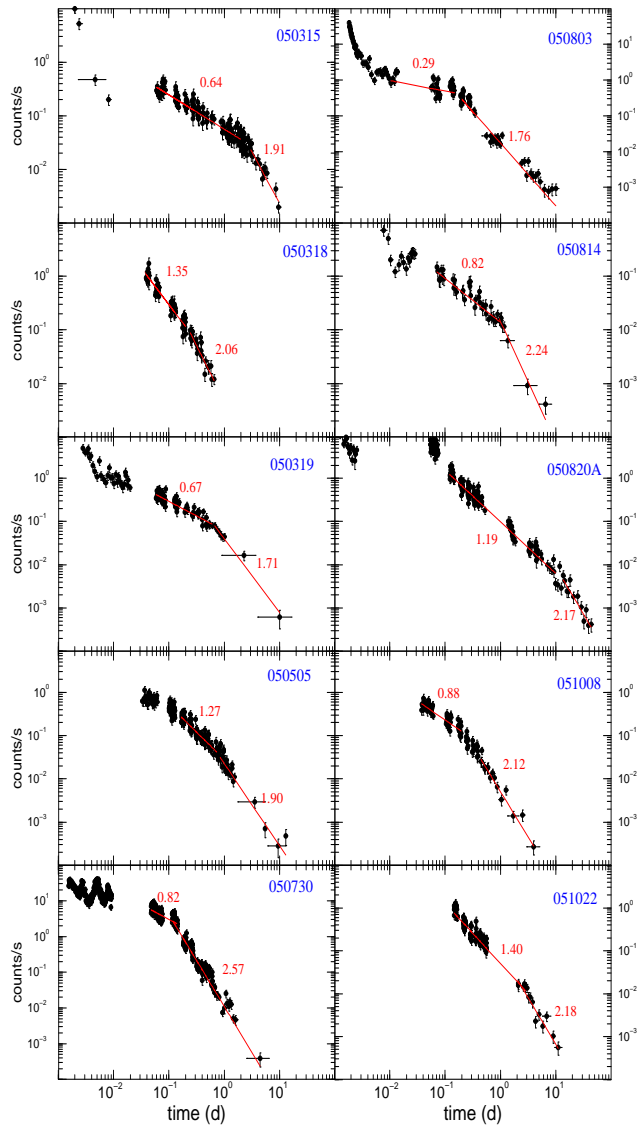,width=8.5cm,height=15cm}}
\caption{ Ten Swift afterglows whose X-ray light-curves exhibit a break at 0.1--10 d
          to a post-break decay faster than $t^{-1.5}$. These steepenings could be
          jet-breaks resulting from the boundary of the GRB outflow becoming visible 
          to the observer. Power-law fits ($C_{0.3-10\,\rm{keV}} \propto t^{-\alpha}$) 
          and the exponents $\alpha$ are shown (standard deviations given in Table 1).}
\label{f1}
\end{figure}

\begin{figure}
\centerline{\psfig{figure=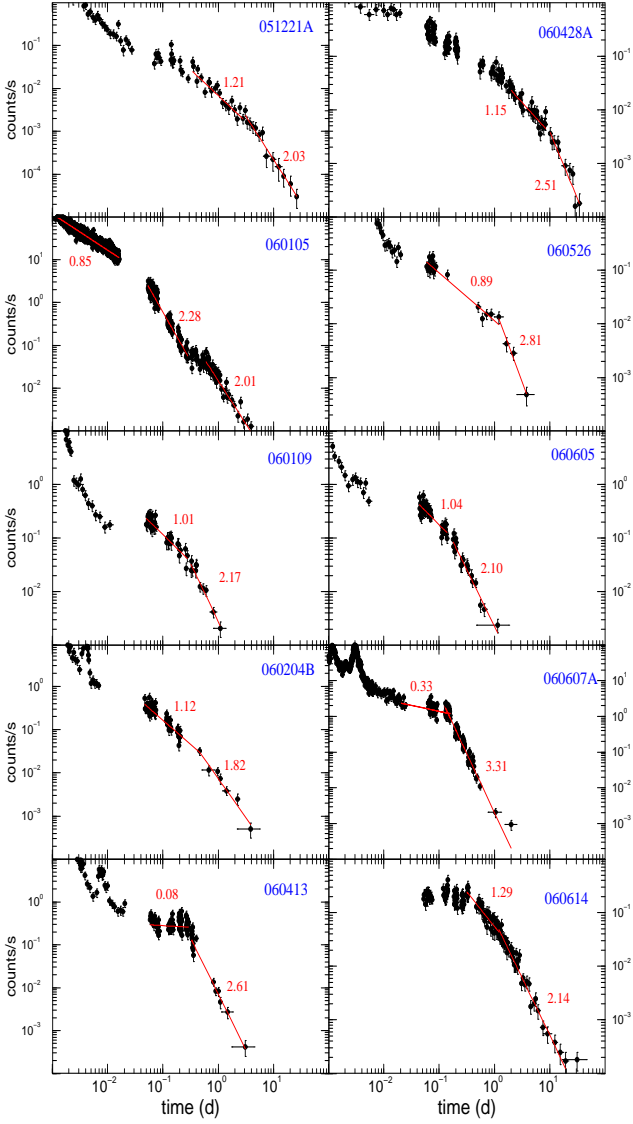,width=8.5cm,height=15cm}}
\caption{ As in Figure \ref{f1}, for another 10 Swift afterglows.}
\label{f2}
\end{figure}

\begin{figure}
\centerline{\psfig{figure=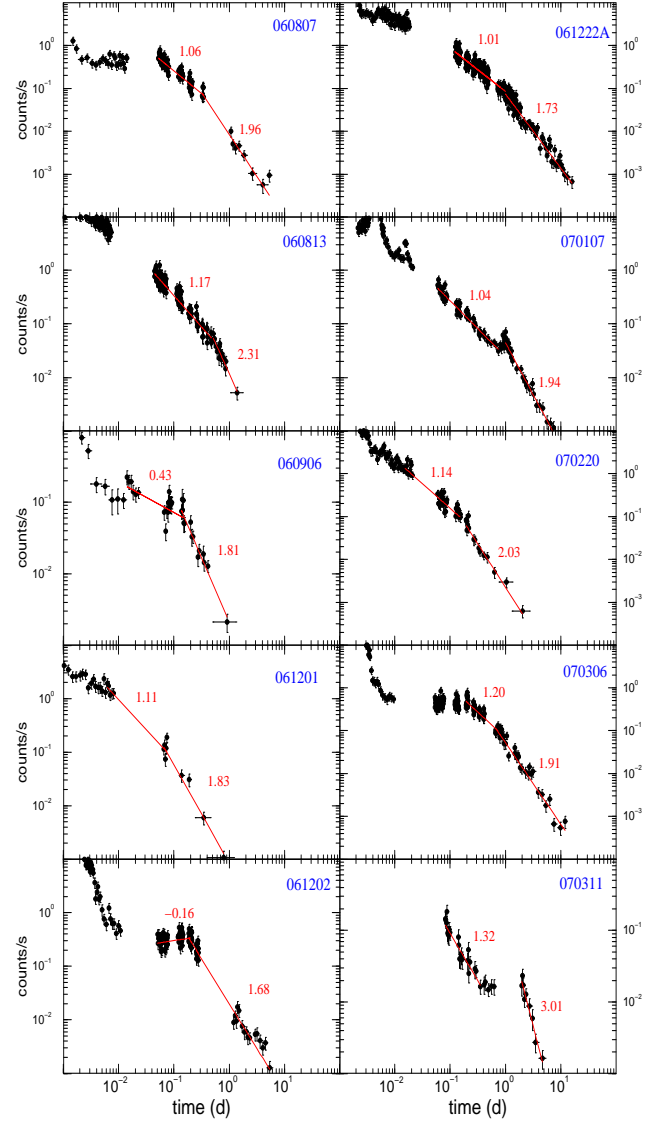,width=8.5cm,height=15cm}}
\caption{ Last set of 10 Swift afterglows with jet-breaks.}
\label{f3}
\end{figure}

 Figures \ref{f1}, \ref{f2}, and \ref{f3} display all afterglows observed by Swift until
April 2007 with good evidence for a late, 0.1--10 day break, followed to a decay steeper 
than $t^{-1.5}$. Their pre and post-break indices ($\alpha_3$ and $\alpha_4$)
of the power-law X-ray light-curve ($C_{0.3-10\,\rm{keV}} \propto t^{-\alpha}$) and the 
slope $\beta$ of the power-law X-ray continuum (spectral distribution of energetic flux 
$F_\nu \propto \nu^{-\beta}$) are listed in Table 1. 
The decay indices were obtained by fitting the 0.3--10 keV count-rate light-curves of
Evans et al (2007); the spectral slopes were obtained through power-law fits in the 1--5
keV range to the specific count-rates $C_\nu \propto \nu^{-(\beta+1)}$ of Butler \& 
Kocevski (2007). 

\begin{figure}
\centerline{\psfig{figure=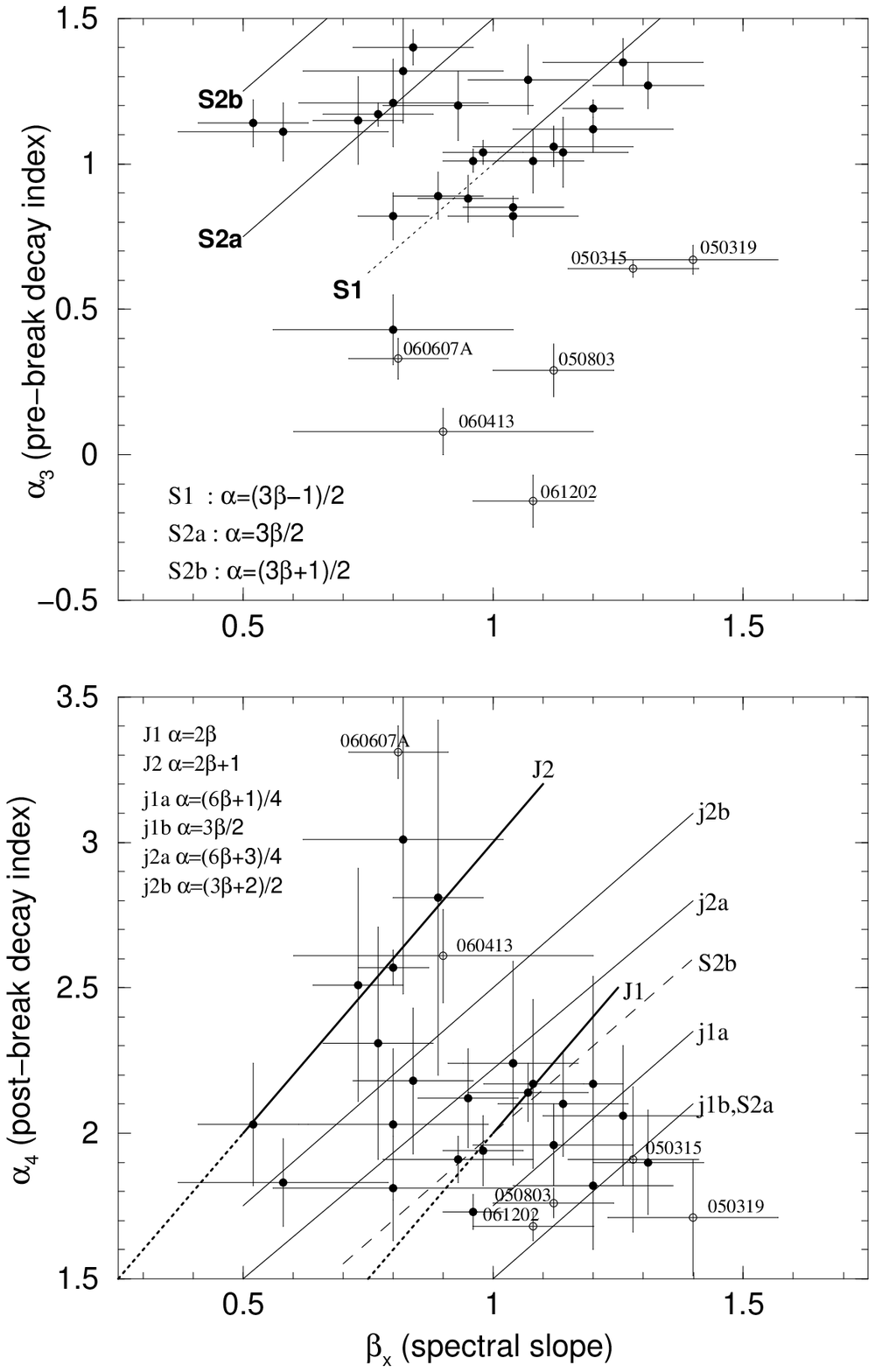,width=8cm}}
\caption{ X-ray spectral slope vs. pre-break (upper panel) and post-break (lower panel)
      decay index for the 30 Swift afterglows with jet-breaks shown in Figures \ref{f1}, 
      \ref{f2} and \ref{f3}. 
     Also shown are the model expectations for the forward-shock synchrotron emission.
     Model labels are as following: 
       "S" = outflow opening wider than the inverse of its Lorentz factor 
            (i.e. a spherical outflow or a jet observed before the jet-break time), 
      "J" and "j" = laterally-spreading and conical jet, respectively, observed 
             after the jet-break time, 
      "1" = cooling frequency ($\nu_c$) below X-ray, "2" = $\nu_c$ above X-ray, 
      "a" = homogeneous medium, "b" = wind-like medium. 
     For "J" and "1" models, the decay index is independent of the radial structure 
      of the ambient medium. 
     Upper panel: six afterglows (labelled) have pre-break decays that are too slow 
         to be explained by the standard forward-shock model.
     Lower panel: one afterglow (060607A) decays too fast after the break to be 
         accommodated by the jet model.  }
\label{ab34}
\end{figure}

 Figure \ref{ab34} compares the decay indices and spectral slopes of these 30 afterglows
with the $\alpha-\beta$ relations expected for the synchrotron emission from the forward-shock
(the standard jet model), assuming that the circumburst medium is either homogeneous or 
has the $r^{-2}$ radial stratification expected for the wind of a massive stellar GRB progenitor. 
The models that reconcile the observed $\alpha$ and $\beta$ are listed in Table 1. 
Derivations of these relations for a spherical outflow and for a spreading jet can be found 
in  M\'esz\'aros \& Rees (1997), Sari, Piran \& Narayan (1998), Chevalier \& Li (1999), 
Rhoads (1999), Sari, Piran \& Halpern (1999), Panaitescu \& Kumar (2000).

 For the post jet-break emission, we also consider the case of a conical jet which does not 
undergo significant spreading, as could happen if the jet edge is not too sharp (Kumar \&
Granot 2003), but surrounded by an envelope which prevents its lateral expansion. 
In this case, the post-break decay index 
$\alpha_4$ is larger by 3/4 (1/2) than the pre-break index $\alpha_3$ for a homogeneous 
(wind-like) medium (Panaitescu, M\'esz\'aros \& Rees 1998), this increase resulting from 
that, after the jet-break, the number of emitting electrons within the "visible" region
of angular extent equal to the inverse of the jet Lorentz factor stops increasing (as the 
jet is decelerated) because the outflow opening is less than that of the cone of relativistic
beaming. Lateral spreading of the jet, if it occurs, enhances the jet deceleration and 
yields an extra contribution to the jet-break steepening $\alpha_4-\alpha_3$ which is smaller 
than the 3/4 (1/2) resulting from the "geometrical" effect described above.

 As can be seen from Figure \ref{ab34} and Table 1, most pre-break X-ray decays require 
either that the cooling frequency ($\nu_c$ is below X-rays (model S1), in which case 
$\alpha_3$ is independent of the ambient medium stratification, or that the medium is 
homogeneous and $\nu_c$ is above X-rays (model S2a). 
 For six afterglows, the pre-break decay is too slow to be explained by the standard 
forward-shock model, indicating a departure from its assumptions. Most likely, this 
departure is the increase of the shock's energy caused by the arrival of fresh ejecta 
(Paczy\'nski 1998, Rees \& M\'esz\'aros 1998, Panaitescu,  M\'esz\'aros \& Rees 1998, 
Nousek et al 2006, Panaitescu et al 2006, Zhang et al 2006). If the post-break decays 
of these six afterglows are attributed to a jet whose boundary is already visible then 
the jet-break time should be before the epoch when energy injection ceases so that, 
shortly after energy injection stops being dynamically important, the jet Lorentz factor 
falls below the inverse of the jet opening. 

 Alternatively, the X-ray light-curve breaks of the six afterglows with slow pre-break
decays could be attributed to the end of energy injection into a spherical (or wide
opening) blast-wave. As shown in the lower panel of Figure \ref{ab34}, the post-break
decays of four of these six afterglows are consistent with the S2a model ($\nu_c$ above
X-ray, homogeneous medium); still, two afterglows (GRBs 060413 and 060607A) exhibit a 
post-break decay that is too steep for any "S" model. 

 Attributing X-ray light-curve breaks to cessation of energy injection can be extended 
to other afterglows: for the measured spectral slope $\beta$, the post-break decays of 19 
afterglows are consistent within $1\sigma$ with that expected for a spherical outflow 
("S" models in the lower panel of Figure \ref{ab34}). 
Here, consistency within $1\sigma$ between a model expectation $\alpha_{model} (\beta) = 
a\beta + b$ and an observed index $\alpha_{obs}$ is defined by $\alpha_{obs}-\alpha_{model}$ 
being within $(\sigma_\alpha^2+a^2\sigma_\beta^2)^{1/2}$ of zero, where $\sigma_\alpha$ and 
$\sigma_\beta$ are the $1\sigma$ standard deviations given in Table 1. 

 Thus only 11 afterglows have a post-break decay that is too fast for a spherical outflow 
and require a jet-interpretation for the break. 
 However, it seems unlikely that cessation of energy injection can be so often the
source of X-ray light-curve breaks, for the following reason. A similar analysis
of the optical decay indices and spectral slopes done for 10 pre-Swift afterglows with 
optical light-curve breaks indicated that six of those breaks are consistent with
arising from an episode of energy injection into a spherical blast-wave, ending at 
the break epoch (Panaitescu 2005a). Nevertheless, the numerical modelling of the radio, 
optical, and X-ray emission of those six afterglows has shown that the broadband 
emission of only two of them can be accommodated by energy injection (Panaitescu 2005b),
the primary reason for the failure of this model being that the radio emission from the 
ejecta electrons that were accelerated by the reverse shock is too bright. 

 Therefore, while it is possible that up to 2/3 of the Swift X-ray breaks listed in Table 1 
arise from cessation of energy injection, this interpretation does not find much support in the
light-curve breaks of pre-Swift afterglows. For this reason, we propose that the 30 Swift X-ray 
breaks of Table 1 are due to the GRB outflow being narrowly collimated, although a different
origin cannot be ruled out based only on X-ray observations. 

 Last column of Table 1 lists the combination of models for the pre- and post-break decays 
that explains both phases in a self-consistent manner, the type of medium and location of 
cooling frequency being the same both before and after the break. The features of those
"global" afterglow models show that:
$(i)$ the cooling frequency must be below X-rays ("1" models) for 2/3 of afterglows,
$(ii)$ 1/4 of afterglows require a homogeneous medium ("a" models), 
$(iii)$ 1/10 require a wind medium ("b" models),
$(iv)$ 1/3 of afterglows require a spreading jet ("J" models), and
$(v)$ 1/3 require a conical jet ("j" models),
larger fractions allowing any of the above features as they are not always well constrained. 
This shows that there is a substantial diversity in the details of the forward-shock model
that accommodate the temporal and spectral properties of Swift X-ray afterglows.

\begin{table*}
 \caption{ Decay indices ($\alpha_3$ before break, $\alpha_4$ after break), 
    spectral slopes ($\beta_x$, stars indicate values taken from Willingale et al 2007), 
    break epoch ($t_b$), and possible models for the 30 Swift X-ray afterglows with jet-breaks. 
    Models consistent with the observations at the $1\sigma-2\sigma$ level are given
    in round parentheses only when no model is consistent within $1\sigma$ with the 
    observed $\alpha$ and $\beta$.
   Model coding is given in captions of Figure \ref{ab34}. "ei" is for energy injection 
      during the pre-break phase. "$\nu_c = \nu_x$" is for cooling frequency crossing
      the X-ray around the jet-break time. }
\begin{tabular}{lcccccllll}
  \hline 
 GRB & $\alpha_3$(sd) & $\beta_x$(sd) & $\alpha_4$(sd) & Pre-break & $t_b$(day) &
      Post-break Model & \hspace{4mm} Global &  & \hspace{-8mm} Model \\
  \hline 
 050315 & 0.64(.03) & 1.28(.13) & 1.91(.25) & ei &  3  & j1a, j1b         & S1+ei&\ra&j1a/j1b   \\
 050318 & 1.35(.08) & 1.26(.16) & 2.06(.24) & S1 & 0.3 & j1a, j1b         & S1&\ra&j1a/j1b      \\
 050319 & 0.67(.05) & 1.40(.17) & 1.71(.20) & ei & 0.7 & (j1a), (j1b)     &S1+ei&\ra&(j1a)/(j1b)\\
 050505 & 1.27(.08) & 1.31(.11) & 1.90(.18) &(S1)& 0.8 & j1b              & (S1)&\ra&j1b        \\
 050730 & 0.82(.08) & 0.80(.07) & 2.57(.06) & S1 & 0.15& J2               & S1 & \xc & J2       \\
 050803 & 0.29(.09) & 1.12(.12) & 1.76(.05) & ei & 0.15& j1a, j1b         & S1+ei&\ra&j1a/j1b   \\
 050814 & 0.81(.07) & 1.04(.13) & 2.24(.35) &(S1)&  1  & J1, j2a, j2b     & (S1)&\ra&J1         \\
 050820A& 1.19(.03) & 1.20(.06) & 2.17(.37) &(S1)& 10  & J1, j1a, j1b, j2a& (S1)&\ra&J1/j1a/j1b \\
 051008 & 0.88(.08) & 0.95(.10) & 2.12(.17) & S1 & 0.25& J1, j2a          & S1&\ra&J1           \\
 051022 & 1.40(.06) & 0.84(.12) & 2.18(.25) & S2a&  3  & j2a, j2b         & S2a&\ra&j2a         \\
 051221A& 1.21(.15) & 0.80(.19) & 2.03(.26) & S2a&  3  & J1, j2a, j2b     & S2a&\ra&j2a         \\
 060105 & 0.85(.01) & 1.04(.10) & 2.24(.06) &(S1)&0.025/0.6& J1, j2a      & (S1)&\ra&J1         \\
 060109 & 1.01(.11) & 1.08(.10)*& 2.17(.29) & S1 & 0.4 & J1, j1a, j2a     & S1&\ra&J1/j1a       \\ 
 060204B& 1.12(.08) & 1.20(.16) & 1.82(.22) & S1 & 0.5 & j1a, j1b         & S1&\ra&j1a          \\
 060413 & 0.08(.08) & 0.9(.3)*  & 2.61(.16) & ei & 0.3 & J2, j2b          & S2a/b+ei&\ra&J2b    \\
  ...   &   ...     &  ...      &  ...      & ...& ... & ...              & S2b+ei&\ra&j2b      \\
 060428A& 1.15(.15) & 0.73(.09) & 2.51(.40) & S2a& 10  & J2, j2b          & S2a&\ra&J2          \\
 060526 & 0.89(.08) & 0.87(.06)*& 0.89(.09) & S1 & 1.5 & J2, j2b          & S1 & \xc & J2       \\
 060605 & 1.04(.12) & 1.14(.13) & 2.10(.18) & S1 & 0.15& J1, j1a          & S1&\ra&J1/j1a       \\
 060607A& 0.33(.07) & 0.81(.10) & 3.31(.09) & ei & 0.15& ???           &  & \hspace{-4mm} ??? & \\
 060614 & 1.29(.12) & 1.07(.12) & 2.14(.10) & S1 & 1.5 & J1               & S1&\ra&J1           \\
 060807 & 1.06(.07) & 1.12(.16) & 1.96(.14) & S1 & 0.3 & J1, j1a          & S1&\ra&J1/j1a       \\
 060813 & 1.17(.04) & 0.77(.11) & 2.31(.40) & S2a& 0.5 & J2, j2a, j2b     & S2a&\ra&J2/j2a      \\
 060906 & 0.43(.12) & 0.80(.24) & 1.81(.18) & S1 & 0.2 & J1, j1a, j2a, j2b& S1&\ra&J1/j1a       \\
 061201 & 1.11(.10) & 0.58(.21) & 1.83(.15) &S2a/S2b&$\siml 0.08$& J2, j2a, j2b & S2a&\ra&J2/j2a\\
 ...    &  ...      &  ...      &  ...      & ...& ... & ...              & S2b&\ra&j2b         \\
 061202 &-0.16(.09) & 1.08(.12) & 1.68(.05) & ei & 0.2 & j1b              & S1+ei&\ra&j1b       \\
 061222A& 1.01(.04) & 0.96(.06) & 1.73(.06) & S1 &  1  & j1a              & S1&\ra&j1a          \\
 070107 & 1.04(.04) & 0.98(.08) & 1.94(.12) & S1 &  1  & J1               & S1&\ra&J1           \\
 070220 & 1.14(.08) & 0.52(.11) & 2.03(.21) & S2b& 0.2 & J2, j2b          & S2b&\ra&J2/j2b      \\
 070306 & 1.20(.12) & 0.93(.15) & 1.91(.08) & S2a& 0.7 & J1, j2a          & S2a&\ra&j2a         \\
 070311 & 1.32(.18) & 0.82(.20) & 3.01(.53) & S2a& 1--2& J2               & S2a&\ra&J2          \\
  \hline 
\end{tabular}
\end{table*}

\section{Swift X-ray and pre-Swift optical breaks}

 In the Jan05--Mar07 set of Swift afterglows, we find another 27 potential jet-breaks at 
0.1--10 d, followed by a $t^{-1.5}$ decay or steeper: GRB 050401, 050408, 050525A, 050603, 
050712, 050713A, 050713B, 050726, 050802, 050826, 050922B, 051001, 051016B, 051211B, 060121, 
060124, 060210, 060218, 060219, 060306, 060707, 060719, 060923C, 061019, 061126, 070125, 070318. 
For some of these afterglows, the post-break decay was followed for only 0.5 dex in time, 
for others, the break is only marginally significant, with the break magnitude ($\alpha_4 - 
\alpha_3$) being smaller than for the 30 afterglows in Table 1. 
Therefore, in the Jan05--Mar07 set of X-ray afterglows, there could be as many as 57 with 
jet-breaks. However, we cannot exclude the possibility that some of those light-curve breaks 
arise from another mechanism, such as the sudden change of energy injected in the forward shock.

 In the same sample, there are 18 afterglows (GRB 050607, 050915B, 051016A, 060108, 060111A, 
060115, 060123,  060510A, 060604, 060708, 060712, 060904A, 060912A, 060923A, 061110A, 070223, 
070224, 070328) exhibiting a $t^{-1.5}$ decay or slower until 3--10 day, indicating 
that their jet-breaks occurred after the last observation. 
Similar decays, but lasting until 10--30 day, are displayed by 13 afterglows (GRB 050716, 
050824, 051021A, 051109A, 051117A, 060202,  060714, 060814, 061007, 061121, 061122, 070110, 
070129) and by 6 afterglows (GRB 050416A, 050822, 060206, 060319, 060729, 061021) until after 
30 day. Thus, the number of well-monitored afterglows without a jet-break is 19, but could be 
as high as 37 if all the other 18 light-curves followed for less than 10 day are included. 

 For the remaining more than 100 afterglows the temporal coverage is insufficient to test for 
the existence of jet-breaks. We conclude that, if only the afterglows with good evidence for 
existence or lack of jet-breaks are counted, then the fraction of Swift afterglows with 
jet-breaks is 30/(30+18)=0.63; if we include all potential cases for each type then that 
fraction is 57/(57+37)=0.61.

 In pre-Swift afterglow observations, evidence for light-curve breaks is found only in 
the optical emission. The X-ray coverage of pre-Swift afterglows at the time of the 
optical break is too limited to test for the existence of a simultaneous break in the 
X-ray emission. 
The radio light-curves of a dozen pre-Swift afterglows show breaks at 1--10 day, however 
the pre and post-break decays indicate that those breaks arose from the passage of the 
synchrotron peak frequency through the radio. All radio post-break decays are slower than 
$t^{-1.5}$, hence a jet origin for those breaks is very unlikely.

 There are 12 pre-Swift optical afterglows with good evidence for a break at 0.3--3 day
to a decay steeper than $t^{-1.5}$: GRB 980519, 990123, 990510, 991216, 000301C, 000926, 
011211, 030226, 030328, 030329, 030429, 041006 (figure 1 of Zeh, Klose \& Kann 2006). 
Three afterglows (GRB 011121, 020124, 020405) may also had a break, though the evidence 
is not so strong. A break to a decay slightly less steep than $t^{-1.5}$ was observed 
for three afterglows (GRB 010222, 020813, 021004). We find only 5 optical afterglows 
followed for more than several days that have a decay slower than $t^{-1.5}$: GRB 970228, 
970508, 980329, 000418, 030323. Therefore the fraction of well-monitored optical afterglows 
with potential jet-breaks is 12/(12+5)=0.71 if only the best cases for light-curve breaks 
are taken into account and 18/(18+5)=0.78 including the other 6 potential optical breaks. 

 Therefore the fraction of Swift X-ray afterglows with light-curve breaks at 0.1--10 day
(60 percent) is slightly smaller than that of pre-Swift optical afterglows with breaks at 
0.3--3 day (75 percent). 

\section{Amati and Ghirlanda relations}
\label{amati}

 Having identified new breaks in afterglow light-curves that may qualify as jet-breaks,
we calculate the jet opening from the break epoch $t_b$, assuming a GRB efficiency of 50 
percent and a homogeneous circumburst medium of particle density $n = 1\; {\rm cm^{-3}}$.
From there, the GRB collimated output is
\begin{equation}
  E_{jet} = 1.9 \times 10^{50} \left( \frac{E_\gamma}{10^{53}\,{\rm erg}} 
          \frac{t_{b,d}}{z+1} \right)^{3/4} \, {\rm erg} \;, 
\label{Ejet}
\end{equation}
where $E_\gamma$ is the burst isotropic-equivalent output in the (host-frame) 1 keV--10 MeV 
range, $t_{b,d}$ is the jet-break epoch measured in days, and numerical coefficient is
for the arrival-time of photons emitted from the jet edge. The results below do not change 
much if a wind-like medium is assumed, for which $E_{jet} \propto (E_\gamma t_b)^{1/2}$,
because we shall use $\log E_{jet}$ in the calculation of the correlation coefficient $r$ and
changing the multiplying factor (3/4 to 1/2) of $\log E_{jet}$ does not affect $r$ (though
it would alter the slope of the best-fits involving  $\log E_{jet}$).
                                                                                                   
\begin{figure*}
\centerline{\psfig{figure=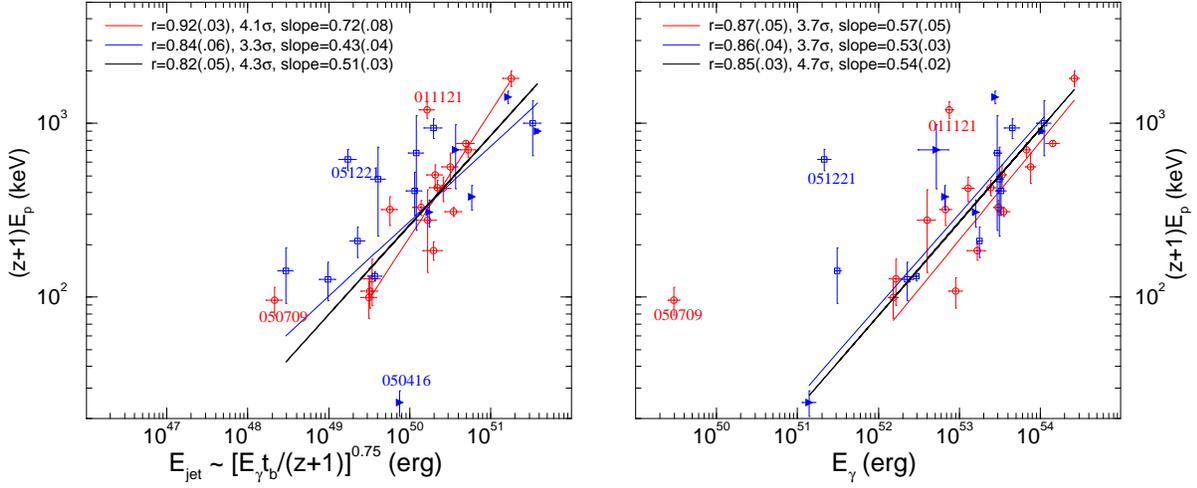,width=16cm}}
\caption{ Correlation of host-frame peak energy of the $\nu F_\nu$ GRB spectrum and 
     $k$-corrected 1 keV--10 MeV collimated GRB output (Ghirlanda relation -- left panel) 
     and isotropic-equivalent GRB energy release (Amati relation -- right panel) for 
     15 pre-Swift afterglows with optical light-curve jet-breaks 
      (red symbols, fit shown with red line), 
     9 Swift afterglows with X-ray jet-breaks (blue squares) and 6 without (blue triangles)
      (fit shown with dashed line), 
     and for the entire set of 30 afterglows (black line). 
     Labelled afterglows are not included. Legends give the linear-correlation coefficient 
     $r$ of plotted quantities in log-log space, significance level for that correlation, 
     and slope of linear best-fit, all accounting for sample variance. } 
\label{ga}
\end{figure*}

 The necessary information (low and high-energy burst spectral slopes, peak energy $E_p$ of the 
$\nu F_\nu$ GRB spectrum, burst fluence and redshift) to study the Amati and Ghirlanda relations
(most recently presented by Amati 2006 and Ghirlanda et al 2007) between the intrinsic peak energy 
$E'_p=(z+1)E_p$ of the burst spectrum and the isotropic GRB output $E_\gamma$ or the collimated GRB 
energy $E_{jet}$ is available for 
\begin{enumerate}
\vspace*{-2mm}
\item 15 pre-Swift optical afterglows: GRB 990123, 990510, 991216, 000926, 010222, 011121, 020124, 
      020405, 020813, 021004, 030226, 030328, 030329, 030429, 041006 
\item 9 of the 57 Swift X-ray afterglows with good or potential evidence for jet-breaks: 
      GRB 050318, 050505, 050525A, 050803, 050814, 050820A, 060124, 060605, 060906
\item 6 of the 37 Swift X-ray afterglows without a jet-break until at least 3 day: 
      GRB 050416A, 051109A,060115, 060206, 061007, 061121
\item 2 short-bursts (lasting less than 2 seconds): GRB 050709 and 051221A. 
\end{enumerate}

 Figure \ref{ga} displays the significance of the Amati and Ghirlanda correlations for the
set 15 pre-Swift afterglows with optical jet-breaks, 15 Swift afterglows with X-ray jet-breaks 
or without one until more than 3 day, and the joint set of 30 afterglows. The long-duration GRB 
011121 and the only two short-bursts 051221A and 050709 were excluded as they are outliers for 
the Amati relation. 
For the Ghirlanda relation, GRB 050416A was excluded as an outlier and the last observation
epoch of the 6 X-ray afterglows without jet-breaks were taken as jet-break times. Their true
jet-break epochs would evidently be later which, as can be seen from Figure \ref{ga}, would
weaken the Ghirlanda relation (see also figure 6 of Sato etal 2007 and figure 9 of Willingale 
et al 2007).

 The linear correlation coefficients and best-fit slopes given in Figure \ref{ga} show that:
\begin{enumerate}
\vspace*{-2mm}
\item Swift X-ray afterglows display the same Amati correlation as the pre-Swift optical 
      afterglows but a weaker Ghirlanda correlation,
\item the addition of Swift afterglows weakens the Ghirlanda correlation (smaller correlation
      coefficient) and increases the statistical significance of the Amati correlation.
      The former result was also pointed out by Campana et al (2007), but we note that half 
      of the 8 X-ray light-curve breaks identified in that work are followed by decays slower 
      than $t^{-1.5}$ and, thus, they may not be jet-breaks,
\item going from isotropic to collimated GRB output brings the three outlying bursts for the
      Amati relation closer to the rest.
\end{enumerate}

 For the entire set of 30 afterglows with optical and/or X-ray light-curve jet-breaks, we find 
a the log-log space slope for the Amati relation ($s_A = 0.54 \pm 0.02$) which is consistent 
with that obtained by Amati (2006) for 41 afterglows ($s_A = 0.49 \pm 0.06$), but a smaller 
one for the Ghirlanda relation ($s_G = 0.51 \pm 0.03$) than that obtained by Ghirlanda et al 
(2007) for 25 afterglows ($s_G = 0.70 \pm 0.04$). 
This discrepancy is due to that we did not include here the afterglows of Ghirlanda (2007) that
have breaks followed by decays shallower than $t^{-1.5}$, which are unlikely to be jet-breaks,
and have used a larger set of Swift afterglows.

 That the $E'_p - E_{jet}$ correlation coefficient ($r_G = 0.82 \pm 0.05$) is nearly the same as 
for $E'_p - E_\gamma$ ($r_A = 0.85 \pm 0.03$) indicates that the addition of a new observable 
(the jet-break time $t_b$) does not reduce the spread of the Amati relation. This suggests that 
the jet-break time is not correlated with either burst observable. Indeed, we find such correlations 
not to be statistically significant: $r(\log E'_p,\log t'_b)=0.25 \pm 0.10$ and $r(\log E_\gamma,
\log t'_b)= 0.14 \pm 0.05$, where $t'_b = t_b/(z+1)$ is the host-frame jet-break time.

 For the 15 pre-Swift afterglows with optical breaks, the slope of the $E'_p - E_{jet}$ best-fit 
($s_G = 0.72 \pm 0.08$) is equal to that of the $E'_p - E_\gamma$ best-fit ($s_A = 0.57 \pm 0.05$) 
multiplied by $\rm{d}\log E_\gamma/\rm{d}\log E_{jet} = 4/3$ (from equation (\ref{Ejet}). 
This indicates that the Ghirlanda relation for pre-Swift afterglows is the consequence of Amati's.
For the entire set of 30 afterglows, $s_G < (4/3) s_A$, consistent with the same conclusion.

\section{Conclusions}

 Summarizing our findings, out of the more than 200 X-ray afterglows monitored by Swift
from January 2005 through March 2007, about 100 have been followed sufficiently long 
to test for the existence of late light-curve steepenings. About 60 percent of these
well-monitored afterglows display a clear or a possible X-ray light-curve break at 0.1--10 
day followed by a $t^{-1.5}$ or steeper decay. These are potential jet-breaks, resulting 
when the jet Lorentz factor decreases below the inverse of the jet opening. However, from 
X-ray observations alone, we cannot exclude other origins for the light-curve breaks.

 More stringent tests of the jet model for X-ray light-curve breaks require a good optical 
coverage. So far, only a couple of the 30 X-ray breaks listed in Table 1 were followed in 
the optical, showing that the X-ray breaks were achromatic, as expected for a jet origin:
GRB 050730 (Pandey et al 2006, Perry et al 2007), GRB 060124 (Curran et al 2007), and
GRB 050526 (Dai et al 2007). For the last two, the pre- and post-break optical and X-ray 
decay indices are consistent with the jet interpretation but, for the first, the post-break
optical light-curve falls-off too slowly ($F_o \propto t^{-1.4\pm0.2}$) compared to the
X-ray emission ($F_o \propto t^{-2.6\pm0.1}$).

 Around 75 percent of the pre-Swift optical afterglows with a good coverage exhibit a 
light-curve break at 0.3--3 day. Hence the fraction of Swift X-ray afterglows with breaks 
is slightly smaller, but comparable, to that of pre-Swift optical afterglows.

 The burst and redshift information necessary to test the Ghirlanda ($E'_p \propto E_{jet}^{0.7}$,
$E'_p$ = peak energy of the burst spectrum, $E_{jet} \propto (E_\gamma t_b)^{3/4}$ = GRB 
collimated output, $E_\gamma$ = GRB isotropic energy release, $t_b$ = afterglow break epoch) 
and Amati ($E'_p \propto E_\gamma^{0.5}$) correlations exist for only eight of the 30 afterglows 
with X-ray jet-breaks. These afterglows display the mentioned correlations at the $\simg 
2\sigma$ level. Adding them to a set of 15 pre-Swift afterglows with optical light-curve breaks, 
leads to correlations which are significant at the $\sim 4\sigma$ level. However, including the 
jet-break time in the Amati correlation does not lead to a stronger (Ghirlanda) correlation, 
unless the few under-energetic outliers shown in Figure \ref{ga} are taken into account. 
Furthermore, because the jet-break time is not correlated with either burst property, the slope 
of the $\log E'_p - \log E_{jet}$ best-fit is that expected from the slope of $\log E'_p - 
\log E_\gamma$ fit. These two facts indicate that the Ghirlanda correlation results almost 
entirely from the Amati correlation.

 For the cosmological use of GRBs, one is interested in obtaining a good calibrator of the source 
luminosity $E_\gamma$ with other burst observables, a quality which is quantified by the linear 
correlation coefficient. 
 Although the Ghirlanda correlation is stronger than Amati's when outliers are included, the 
Ghirlanda relation is not necessarily better for constraining cosmological parameters because 
outliers to the Amati relation, such as the three underluminous bursts shown in Figure \ref{ga}, 
should stand out and be easily to excise from the Hubble diagram constructed based on the Amati 
relation (i.e. with the luminosity distance inferred from the burst isotropic luminosity obtained 
from the peak energy of the burst spectrum). 
 As shown in \S\ref{amati}, if we consider only bursts which are not outliers to the Amati relation, 
the addition of a new observable (the afterglow jet-break epoch) does not yield a stronger correlation 
than that of Amati's. 
 Thus, with the current sample of afterglows with jet-breaks, we suggest that the Amati relation 
should be at least as useful for constraining cosmological parameters as is the Ghirlanda relation
(Ghirlanda et al 2004, Schaefer 2007).

\section*{Acknowledgments}
 This work made use of data supplied by the UK Swift Science Data Center at the University 
 of Leicester.

\end{document}